\documentclass[epj]{svjour}
\usepackage{epsfig}
\begin{document}
\title{On the anomalous large-scale flows in the Universe}
\author{Davor Palle}
\institute{
Zavod za teorijsku fiziku, Institut Rugjer Bo\v skovi\' c \\
Bijeni\v cka cesta 54, 10002 Zagreb, Croatia}

\date{Revised: 19 June 2010}

\abstract
{Recent combined analyses of the CMB and galaxy cluster data
reveal unexpectedly large and anisotropic peculiar velocity
fields at large scales. We study cosmic models with included
vorticity, acceleration and total angular momentum of the Universe
in order to understand the phenomenon. The Zel'dovich model is used
to mimic the low redshift evolution of the angular momentum.
Solving coupled evolution equations of the second order
for density contrast in corrected Ellis-Bruni covariant
and gauge-invariant formalism one can properly normalize and
evaluate integrated Sachs-Wolfe effect and peculiar velocity
field. The theoretical results compared to the observations
favor a much larger matter content of the Universe
than that of the concordance model. Large-scale flows appear
anisotropic with dominant components
placed in the plane perpendicular to the axis of
vorticity (rotation). The integrated Sachs-Wolfe term has
a negative contribution to the CMB fluctuations
for the negative cosmological constant and it can 
explain the observed small power of the CMB TT
spectrum at large scales.
The rate of the expansion of the Universe may be substantially affected 
by the angular momentum if its magnitude is large enough. 
}

\maketitle

\section{Introduction and motivation}

Modern cosmology relies heavily on two paradigms in order to
fit vast astrophysical datasets: 1. the existence of dark matter and
2. the existence of dark energy.
It is assumed that dark matter consists of the neutral weakly interacting
fermion decoupled from primordial plasma as a nonrelativistic
species (cold dark matter=CDM). The CDM-mass-density behavior on
redshift is the standard one of nonrelativistic matter.
The origin and the redshift behavior of dark energy is rather obscure.
The concordance $\Lambda CDM$ model presumes that dark energy is positive
cosmological constant.

Dark matter is thus a problem of particle physics, while
dark energy is a common problem of field theory, particle physics and
the theory of gravity.

The common mathematical difficulties of the zero distance singularity 
and causality motivate the proposal for a new symmetry-breaking 
mechanism in particle physics (BY theory of ref. \cite{Palle1}) 
and new studies in the Einstein-Cartan theory of gravity.
Heavy Majorana
neutrinos within the nonsingular and causal SU(3) electroweak-strong
unification \cite{Palle1} represent the main building block
of the Universe as a cold dark matter (DM) particles. They are 
cosmologically stable $\tau_{N_{i}} >> \tau_{U}$ with
large annihilation cross sections \cite{Palle2}. The H.E.S.S. source
J1745-290 \cite{HESS} in the center of our galaxy and WMAP haze
\cite{Finkbeiner} are
possible indications of the heavy DM particle annihilations.

Light Majorana neutrinos trigger vorticity of the Universe
\cite{Palle3,Palle4} through the spin-torsion relations in the 
Einstein-Cartan cosmology at early times of the evolution.
The formation of large-scale structures in the form of
galaxies and clusters, as well as, the present anisotropy of spacetime,
not predicted by inflation of the concordance $\Lambda CDM$ model,
indicate the existence of nonvanishing total 
angular momentum of the Universe.

The current measurements of the CMB by WMAP, the large catalogues of SDSS,
the new cluster and the peculiar velocity catalogues motivate
us to undertake considerations to explain features that are
not expected in the concordance $\Lambda CDM$ model-anisotropic 
anomalous large-scale flows.
The low power of large-scale TT CMB fluctuations observed by
COBE and WMAP, discrepant with the $\Lambda CDM$ model, should be
explained with new cosmologies.

In the next chapter we derive the evolution equations of
density contrast for models with expansion, vorticity,
acceleration and angular momentum within the corrected
Ellis-Bruni formalism. A comparison is made with the 
standard formulas of isotropic and homogeneous spacetime.
One can then evaluate integrated Sachs-Wolfe effect
and RMS peculiar velocity field for any model.

The last chapter is devoted for discussions, conclusions
and suggestions for future work.
Appendices contain detailed framework for Einstein-Cartan
cosmology with field equations, propagation equations and a comparison
of the corrected and standard Ellis-Bruni fluid-flow approach for
the density-contrast evolution. 

\section{Evolution equations}

Our interest is the evolution of the Universe during the matter
dominated epoch. The standard lore for the evolution
of the density contrast, peculiar acceleration and
peculiar velocity field gives the following equations
\cite{Peebles,Kolb,Paddy}:

\begin{eqnarray*}
\delta (a,\vec{k})\equiv \delta(a)\delta_{\vec{k}},\ 
\delta (\vec{x}) =\int d^{3}k e^{\imath \vec{k}\cdot\vec{x}}
\delta_{\vec{k}},
\end{eqnarray*}
\begin{eqnarray}
\vec{v}_{\vec{k}}=\frac{\imath \vec{k}}{k^{2}} a \dot{a}
\delta_{\vec{k}} \frac{d \delta (a)}{d a},\ 
a = R(t)/R_{0}=1/(1+z).
\end{eqnarray}

The root mean square (RMS) values of the mass-density contrast and the peculiar
velocity field at a certain scale $S$ and the redshift $z$ using 
Gaussian window functions \cite{Peebles,Kolb,Paddy}
are defined as:

\begin{eqnarray}
(\delta M/M)^{2}_{RMS}(a,S) \equiv
\langle (\delta M/M)^{2} \rangle (a,S)= N V^{-2}_{W}\int d^{3}k W^{2}(\vec{k},S)
\delta^{2}(a) |\delta_{\vec{k}}|^{2}, 
\end{eqnarray}
\begin{eqnarray}
v_{RMS}^{2}(a,S) \equiv \langle v^{2} \rangle (a,S)=N V^{-2}_{W}
\int d^{3}k W^{2}(\vec{k},S) \frac{1}{\vec{k}^2}
(a \dot{a}\frac{d \delta (a)}{d a})^{2} |\delta_{\vec{k}}|^{2}, \\
V_{W}=\frac{4 \pi}{3} S^{3},\ W(\vec{k},S)=(2\pi)^{3/2}S^{3}
e^{-\frac{1}{2} S^{2}\vec{k}^{2}}. \nonumber
\end{eqnarray}

The aim is to study cosmological models with vorticity, acceleration
and nonvanishing total angular momentum through torsion effects.
We use the standard CDM power spectrum $P(k)=|\delta_{\vec{k}}|^{2}$
defined in \cite{Kolb}.

The growth function $\delta (a)$ must be studied carefully.
Our choice of covariant spacelike vectors within a fluid-flow
formalism differs from that of Ellis and Bruni \cite{Ellis}:

\begin{eqnarray}
\delta=(-{\cal D}_{\mu}{\cal D}^{\mu})^{1/2},\ 
{\cal D}_{\mu} &\equiv & R^{2}(t)\rho^{-1}h_{\mu}^{\ \nu}
\tilde{\nabla}_{\nu} \rho, \\
{\cal L}_{\mu} &\equiv &R^{2}(t)h_{\mu}^{\ \nu}
\tilde{\nabla}_{\nu} \Theta. \nonumber
\end{eqnarray}

These vectors fulfil the Stewart-Walker lemma \cite{Stewart} and
their evolution equations result in a correct solution
for a density contrast 
formed from their scalar invariants. This is not the case
for the standard Ellis-Bruni covariant vectors. 

The detailed derivation of the equations one can find in 
Appendix A while a comparison between two fluid-flow approaches is in 
Appendix B.

The resulting second order coupled equations in our corrected
scheme are given by (matter dominated regime):

\begin{eqnarray}
\ddot{{\cal D}}_{\mu}-\frac{1}{3}\dot{\Theta}{\cal D}_{\mu}
-\frac{1}{3}\Theta \dot{{\cal D}}_{\mu}+a_{\mu}a^{\nu}{\cal D}_{\nu}
+u_{\mu}\dot{a}^{\nu}{\cal D}_{\nu}
+u_{\mu}a^{\nu}\dot{{\cal D}}_{\nu}  \nonumber \\
-(\frac{1}{3}\Theta \delta^{\lambda}_{\mu}+u_{\mu}a^{\lambda}
+\tilde{\omega}^{\lambda}_{.\mu}+\sigma^{\lambda}_{.\mu})
(\frac{1}{3}\Theta{\cal D}_{\lambda}-u_{\lambda}a^{\nu}{\cal D}_{\nu}
-\dot{{\cal D}}_{\lambda}-{\cal D}_{\rho}(\sigma^{\rho}_{.\lambda}+
\tilde{\omega}^{\rho}_{.\lambda})+\Theta R^{2}a_{\lambda})  \nonumber \\
-2 a_{\mu} R^{2} \dot{\Theta}-R^{2} (^{(3)}\tilde{\nabla}_{\mu} {\cal N})
-\frac{1}{2}\kappa \rho {\cal D}_{\mu}
+\dot{{\cal D}}_{\lambda} (\sigma^{\lambda}_{.\mu}+
\tilde{\omega}^{\lambda}_{.\mu})  \nonumber \\
+{\cal D}_{\lambda} (\dot{\sigma}^{\lambda}_{.\mu}
+\dot{\tilde{\omega}}^{\lambda}_{.\mu})
-\frac{2}{3} R^{2}\Theta^{2}a_{\mu}-\Theta R^{2}\dot{a_{\mu}}=0,
\end{eqnarray}
\begin{eqnarray*}
{\cal N}=2\sigma^{2}-2\tilde{\omega}^{2}-\tilde{\nabla}_{\mu}a^{\mu}.
\end{eqnarray*}

We use the Zel'dovich model to describe the evolution of the total
angular momentum of the Universe at small redshifts 
\cite{Paddy} assuming the surplus of right-handed over
left-handed galaxies and clusters \cite{Palle4,Longo}

\begin{eqnarray}
\vec{L}(t)\propto a^{2}\dot{a}\bar{\rho}\int_{V_{L}}
d^{3}q (\vec{q}-\vec{\bar{q}})\times [\vec{\nabla}\Phi_{0}(q)
-\vec{\nabla}\Phi_{0}(\bar{q})],  \nonumber \\
\vec{L}(Universe) \propto [n(right)-n(left)]\vec{L}(t), \nonumber \\
Q=torsion \propto L(Universe) \Rightarrow Q(a)=Q_{0}a^{-3/2}\ for\ 
z=a^{-1}-1 < z_{cr}=4, \\
 where\ z_{cr}=4\ is\ put\ arbitrarily,\ otherwise\ Q(z > z_{cr})=0.
\nonumber
\end{eqnarray}

The Einstein-Cartan equations remain unaltered with respect to the
functional form of the time dependence
of torsion (see Appendix A).

We have to factorize density contrast on the space and time dependent
parts. One can achieve this goal transforming the evolution equations
for covariant vectors to the local Lorentzian frame by
tetrads:

\begin{eqnarray*}
{\cal D}_{a} = v^{\mu}_{a}{\cal D}_{\mu},\ 
g^{\mu\nu} = v^{\mu}_{a}v^{\nu}_{b}\eta^{a b},\ 
\delta =(-{\cal D}_{a}{\cal D}^{a})^{1/2},
\end{eqnarray*}
\begin{eqnarray*}
\eta^{a b}=diag(+1,-1,-1,-1),\ \mu,\nu=0,1,2,3,
\ a,b=\hat{0},\hat{1},\hat{2},\hat{3}. 
\end{eqnarray*}

In the Appendix A one can find evaluated coefficients for
the following spacetime metric:

\begin{eqnarray}
d s^{2}=d t^{2}-R^{2}(t)[d x^{2}+(1-\lambda^{2}(t))e^{2mx}d y^{2}]
-R^{2}(t)d z^{2}-2 R(t)\lambda(t)e^{mx}d y dt, \\
m=const. \nonumber
\end{eqnarray}

It is easy to verify that in the Friedmann limit one recovers
the standard form of the density contrast \cite{Cooray}
(see Appendix A):

\begin{eqnarray}
\delta (a) = \frac{H(a)}{H_{0}}\int ^{a}_{0}
d a a^{-3} [\frac{H_{0}}{H(a)}]^{3}, \\
H(a)=H_{0} (\Omega_{m}a^{-3}+\Omega_{\Lambda})^{1/2}. \nonumber
\end{eqnarray}

One can evaluate properly normalized peculiar velocities
and integrated Sachs-Wolfe effect for various cosmological
models \cite{Cooray}:

\begin{eqnarray}
a_{lm}^{ISW} = 12\pi \imath^{l}\int d^{3}k Y^{m*}_{l}(\hat{k})
\delta_{\vec{k}} (\frac{H_{0}}{k})^{2}
\int da j_{l}(kr)\chi^{ISW}, \\
\chi^{ISW}=-\Omega_{m}\frac{d}{da} (\delta (a)/a),
\ r=\int^{1}_{a} da a^{-2}H^{-1}(a),\ \delta (a=1)=1. \nonumber
\end{eqnarray}

\section{Results and discussion}

Supplied with all the necessary equations we can evaluate
properly normalized peculiar velocities and integrated
Sachs-Wolfe effect.
We use the same normalization for all cosmological
models:

\begin{eqnarray*}
(\delta M/M)_{RMS}(a=1,S=10 h^{-1}Mpc) = 1 .
\end{eqnarray*}

This is the standard normalization suitable for the study of 
peculiar velocities (see p.262 of ref.\cite{Paddy}).
We aim to compare cosmic models with respect to the concordance
$\Lambda CDM$ model, not to perform a precision fit to data.

Let us fix relevant parameters of the models in Table I
(unit $H_{0}=1$ is used for parameters m, $\lambda_{0}$ and $Q_{0}$;
EdS=Einstein-de Sitter, EC=Einstein-Cartan,
$\Lambda CDM$=concordance model, $\Omega_{\Lambda}=1-\Omega_{m}$,
 $\lambda=\lambda_{0}R^{-1}$, $Q=Q_{0}R^{-3/2}$, $R_{0}=H_{0}^{-1}$ ). 

\begin{table}
\caption{Model parameters}
\hspace{60 mm} \begin{tabular}{| c || c | c | c | c | c |} \hline
  & $\Omega_{m}$ & h & m & $\lambda_{0}$ & $Q_{0}$ \\  \hline \hline
$\Lambda$CDM & 0.3 & 0.7 & 0 & 0 & 0  \\  \hline 
EdS  &  1  & 0.5  &  0  &  0  &  0   \\  \hline
EC1  &  2  &  0.4 & 0.03 & 0.067 & -0.2  \\  \hline
EC2  &  2  &  0.4 & 0 & 0 & 0 \\  \hline
EC3  &  2  &  0.4 & 0.3 & 0.67 &  0 \\ \hline
\end{tabular}
\end{table}

The formation of small-scale structures and the age of the Universe
can be explained with a larger mass-density and smaller Hubble
constant (see Table II).
This statement is valid if we assume
that the total angular momentum of the Universe at low redshifts, acting
through torsion terms, is much smaller than mass-density
terms. The evolution with large torsion terms must contain
feedback from matter to the background geometry, changing
substantially its expansion and vorticity.
It is possible to utilize this approach within N-body simulations.
Thus, we limit numerical evaluations in this paper to small torsion
contributions.

\begin{table}
\caption{Age of the Universe}
\hspace{60 mm} \begin{tabular}{| c || c | c | c | c |} \hline
  & $\Lambda$CDM & EdS & EC1 & EC2  \\  \hline \hline
$\tau_{U}$(Gyr)  &  13.77  &  13.33 & 13.14 & 13.09  \\ \hline
\end{tabular}
\end{table}

From Tables III-VI it is clear that only models with large mass-density
can enhance large-scale peculiar velocities observed in the analyses
with combined cluster and WMAP data \cite{Kash,Watk,Lavaux}.
Rather small amounts of vorticity, acceleration or torsion do
not essentially influence RMS velocities. However, components
of a density contrast ${\cal D}_{\hat{i}},i=1,2,3$ are not equal.
We assume the standard scaling of the vorticity \cite{Peebles}:
$\lambda(a)=\lambda_{0}a^{-1} \Rightarrow \omega \propto a^{-2}$
(see Apendix for definitions).

\begin{table}
\caption{$v_{rms}(km/s)$ for  the $\Lambda$CDM model}
 \hspace{50 mm} \begin{tabular}{| c || c | c | c | c | c |} \hline
 z $\backslash$ S(Mpc) & 50 & 187.5 & 325 & 462.5 & 600  \\  \hline \hline
 0  & 511.67   & 184.73   & 111.75  & 79.95   &  62.21  \\ \hline
 0.25 & 515.49   & 186.11   & 112.58  & 80.55  &  62.67  \\ \hline
 1  & 467.16   & 168.66   & 102.03  &  73.00 &  56.80  \\ \hline
\end{tabular}
\caption{$v_{rms}(km/s)$ for the EdS model}
\hspace{50 mm} \begin{tabular}{| c || c | c | c | c | c |} \hline
 z $\backslash$ S(Mpc) & 50 & 187.5 & 325 & 462.5 & 600  \\  \hline \hline
 0  & 766.27   & 245.82   & 145.55  & 103.27   &  80.00  \\ \hline
 0.25 & 685.38   & 218.87   & 130.18  & 92.37  &  71.56  \\ \hline
 1  & 541.84   & 173.82   & 102.92  &  73.03 &  56.57  \\ \hline
\end{tabular}
\caption{$v_{rms}(km/s)$ for the EC1 model}
\hspace{50 mm} \begin{tabular}{| c || c | c | c | c | c |} \hline
 z $\backslash$ S(Mpc) & 50 & 187.5 & 325 & 462.5 & 600  \\  \hline \hline
 0  & 1090.94   & 335.81   & 197.54  & 139.83   &  108.19  \\ \hline
 0.25 & 901.75   & 277.58   & 163.28  & 115.58  &  89.42  \\ \hline
 1  & 678.04   & 208.71   & 122.78  &  86.91 &  67.24  \\ \hline
\end{tabular}
\caption{$v_{rms}(km/s)$ for the EC2 model}
\hspace{50 mm} \begin{tabular}{| c || c | c | c | c | c |} \hline
 z $\backslash$ S(Mpc) & 50 & 187.5 & 325 & 462.5 & 600  \\  \hline \hline
 0  & 1091.69   & 336.04   & 197.68  & 139.92   &  108.26  \\ \hline
 0.25 & 902.28   & 277.74   & 163.38  & 115.65  &  89.48  \\ \hline
 1  & 678.45   & 208.84   & 122.85  &  86.96 &  67.28  \\ \hline
\end{tabular}
\end{table}

The density contrasts normalized at zero-redshift do not depend 
on the initial cosmic scale factor, but 
a difference between components does depend.
One can estimate the resulting angle between the axis of vorticity
(z-axis) and the anisotropic bulk velocity. The angle depends on
the initial redshift and the magnitude of the vorticity
($\omega(t)=\frac{1}{2}m\lambda (t)R(t)^{-1}$):

\begin{eqnarray*}
a(initial) &=& 10^{-2},\ a(final)=1,\ model=EC1,\ \frac{\omega_{0}}
{H_{0}}=\frac{1}{2}m\lambda_{0}=10^{-3}  \\
\Rightarrow & &
\angle (\hat{n}(flow),\hat{n}(axis)) = arctg\frac{({\cal D}_{\hat{1}}^{2}
+{\cal D}_{\hat{2}}^{2})^{1/2}}{|{\cal D}_{\hat{3}}|}=
55.3^{\circ}, 
\end{eqnarray*}
\begin{eqnarray*}
a(initial) &=& 10^{-2},\ a(final)=1,\ model=EC1,\ but\ m=0.15:\ 
\angle (\hat{n}(flow),\hat{n}(axis))=57.25^{\circ}, \\
a(initial) &=& 10^{-3},\ a(final)=1,\ model=EC1:\ 
\angle (\hat{n}(flow),\hat{n}(axis))=72.3^{\circ}.
\end{eqnarray*}

Since the metric describes rotation around z-axis, the dominant
components of over(under)densities, the
peculiar accelerations and the velocities are placed in the plain
perpendicular to the axis of rotation (vorticity).
We estimate the angle between the
measured directions of the axis of vorticity \cite{Mague}  and 
the large-scale flows
 \cite{Watk}:

\begin{eqnarray*}
\hat{n}(flow)=(l=287^{\circ} ,b=8^{\circ}),\ 
\hat{n}(axis)=(l=260 ^{\circ},b=60^{\circ})\\
\Rightarrow \angle (\hat{n}(flow),\hat{n}(axis))=53^{\circ}.
\end{eqnarray*}

The reader can compare and visualize
density contrasts and their derivatives for two crucial models ($\Lambda$CDM 
and EC) in Figs. 1 and 2.

\begin{figure}
\epsfig{figure=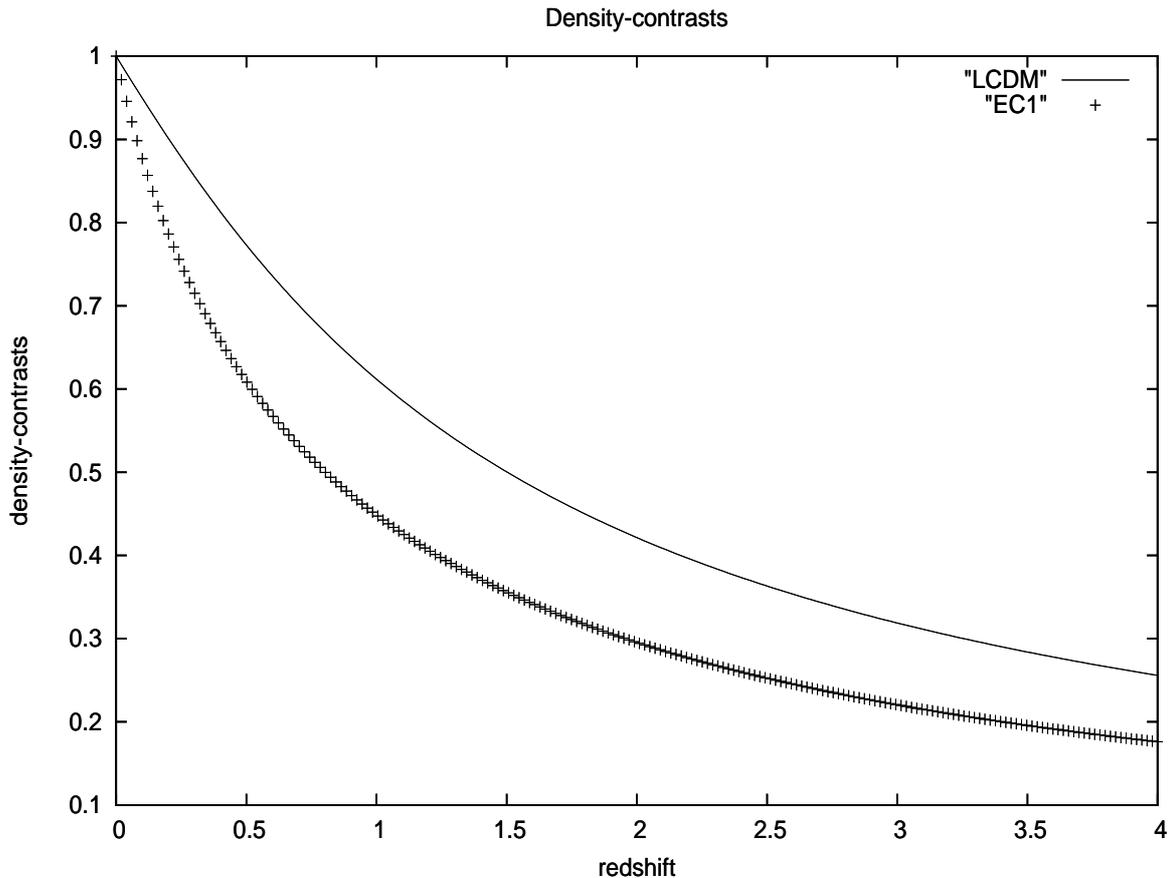, height=120 mm, width=160 mm}
\caption{Density contrasts for the $\Lambda$CDM and EC1 models.}
\end{figure}

\begin{figure}
\epsfig{figure=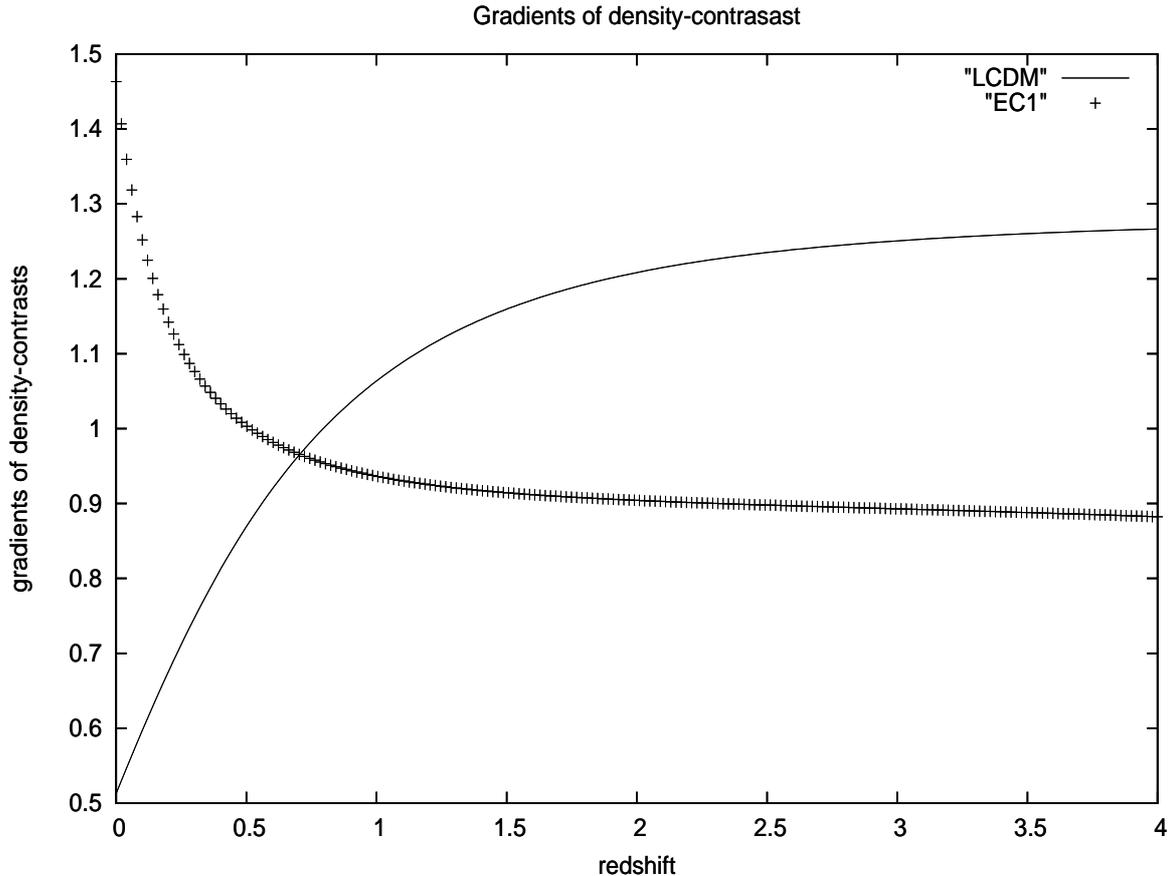, height=120 mm, width=160 mm}
\caption{Gradients of the density contrasts $\frac{d \delta}{d a}$ 
for the $\Lambda$CDM and EC1 models.}
\end{figure}

The integrated Sachs-Wolfe effect is negative for large mass-density
models (EC) with negative cosmological constant, while it is positive
for $\Lambda$CDM, as can be seen in Fig. 3. The negative contribution
of the ISW decreases the total amplitude of the CMB fluctuation, 
while the positive ISW of the $\Lambda$CDM increases it \cite{WMAP}.
The observations point to very small power at large scales,
in contradiction with the $\Lambda$CDM model.

\begin{figure}
\epsfig{figure=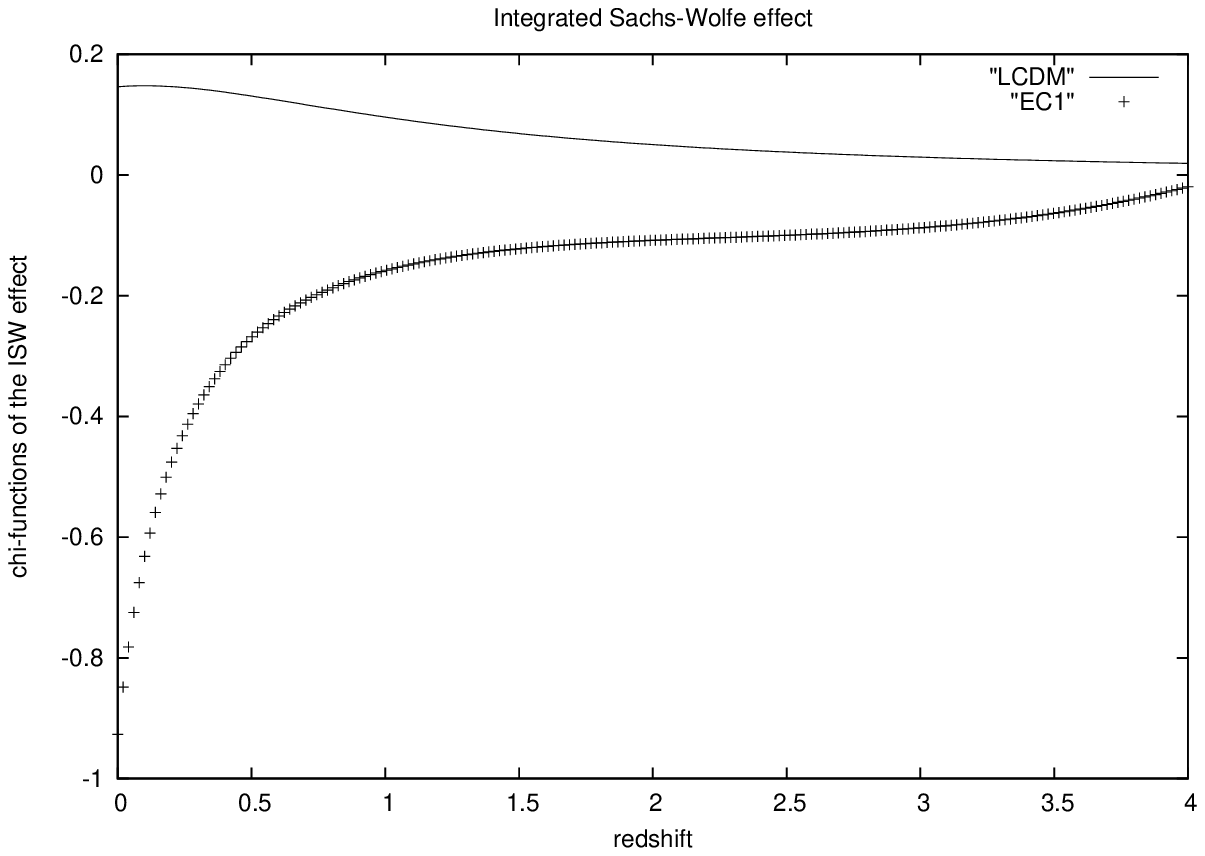, height=120 mm, width=160 mm}
\caption{Integrated Sachs-Wolfe $\chi^{ISW}$ function for the $\Lambda$CDM and
EC1 models.}
\end{figure}

It seems that the introduction of rotational degrees of freedom
(torsion, spin, vorticity, angular momentum) is inevitable in order
to understand
and fit all observational data. Two scenarios emerge as
viable resolutions: (1) small Hubble constant with small amount of 
the total angular momentum of the Universe at present or
possibly: (2) larger Hubble constant if the total angular momentum 
appears much larger. Torsion terms (linear and quadratic) always
give a negative contribution to the effective mass-density, as it
can be seen from Einstein-Cartan field equations (see Appendix A).

Recent measurements of the Hubble constant \cite{Riess} give a
large value, thus EC1 type models with small contribution of
torsion (angular momentum) at low redshifts are ruled out.
The concordance model cannot accommodate to the low power of density fluctuations
at large scales because of the positive contribution of the ISW effect for 
the positive cosmological constant. It has unsurmountable difficulties 
to explain large peculiar velocities, while the observed anisotropies
of the CMB fluctuations and peculiar velocities are complete surprise
for the astrophysical community violating fundamental cosmological
principle of the isotropic Universe.

Let us briefly comment some results of the analyses of cluster data
by various projects and groups. Some groups conclude that the mass
density of the Universe is low like in the concordance model 
\cite{Lima}, while other groups draw a conclusion for a large
mass density \cite{Vauclair}. Large uncertainty in the L-T relations
is a probable cause of the discrepant conclusions.

There is a general belief that the CMB measurements are the most
reliable tool to constrain cosmic parameters. One cannot ignore
disturbing objections for possibly erroneous analyses of time
ordered data \cite{Liu} or problems with beam profile sensitivity
\cite{Sawangwit} in WMAP papers. Any error can affect the estimate
of cosmic parameters.

The equation of state for dark energy is a target of an extensive
research by supernovae \cite{Kessler} or by combined data \cite{Basil}.
Even tests of the concordance model by studying local group 
dynamics provoke more questions than answers \cite{Kroupa}.
It should be mentioned that the unusual time dilation in quasar
light curves \cite{Hawkins} should be eventually explained by
lensing within different cosmic models.

The second scenario with large torsion content at low redshifts
is the most plausible model capable to surmount difficulties
of the concordance model and EC1 type models.
From the work in ref.\cite{Palle3} one can conclude that
$\lim_{R \rightarrow \infty}\rho_{M}/\rho_{\Lambda}=-2$,
but $\lim_{R \rightarrow \infty}\rho_{M}=0$ and then
$\rho_{\Lambda}=0$. It follows that the torsion contribution
plays a role of the negative dark energy (see \cite{Palle3} and Eq.(16))
$\rho_{M,0}/\rho_{torsion,0} \simeq -2$,
making possible the introduction of the large Hubble parameter necessary
for more accurate cosmic clocks and the age of the Universe
(see Eq.(17)). 
Considerations with large angular momentum (torsion) of the Universe
must include N-body numerical simulations.
Dark energy, described by a torsion, should be a clustered
physical quantity \cite{Basilakos} dependent on redshift \cite{Bean}.

To conclude, it is of great importance to search for an independent
method to fix $\rho_{M}$, not only the total density $\rho_{tot}$.
The idea of Zwicky \cite{Zwicky} to study galaxy and cluster
catalogues (SDSS etc.) to estimate directly the mass density of matter
could be advantageous.

\section{Appendix A}

We provide here a complete set of conventions, identities and
equations for Einstein-Cartan theory.
$g_{\mu\nu}$ ia a metric tensor defined in Eq.(7), R(t) is a cosmic
scale factor, $\lambda$(t) function is proportional to the
acceleration vector $a_{\mu}$ in Eq.(11),
m parameter of $g_{\mu\nu}$ defines vorticity in Eq.(11),
$Q^{\alpha}_{.\ \beta\gamma}$ is a torsion tensor, $S^{\alpha}_{.\ \beta\gamma}$
is a spin tensor, $v^{\mu}_{a}$ are tetrad fields, $\rho$ denotes
mass density, p denotes pressure of the fluid, $\Lambda$ is a cosmological
constant, $u_{\mu}$ is a velocity four-vector, $Q^{2}=1/2 Q_{\mu\nu}Q^{\mu\nu},
\ Q^{\alpha}_{.\ \mu\nu}=u^{\alpha}Q_{\mu\nu}$,
$a=R(t)/R_{0}$ in Eq.(17) is a dimensionless cosmic scale factor,
h in Eq.(17) is a Hubble parameter, A in Eq.(18) is a
normalization constant.

The metric is defined as:

\begin{eqnarray*}
g^{\mu\nu} = v^{\mu}_{a}v^{\nu}_{b}\eta^{a b},\ 
\eta_{a b}=diag(+1,-1,-1,-1),\\  
 \mu,\nu=0,1,2,3, \ a,b=\hat{0},\hat{1},\hat{2},\hat{3}, \\
\tilde{\Gamma}^{\alpha}_{\beta\mu}=\Gamma^{\alpha}_{\beta\mu}
+Q^{\alpha}_{. \beta\mu}+Q_{\beta\mu .}^{\ .  . \ \alpha}+
Q_{\mu\beta .}^{\ .  .  \ \alpha},
\end{eqnarray*}
\begin{eqnarray*}
\tilde{R}^{\lambda}_{. \sigma\mu\nu} = 
\partial_{\mu} \tilde{\Gamma}^{\lambda}_{\sigma\nu}
-\partial_{\nu} \tilde{\Gamma}^{\lambda}_{\sigma\mu}
+\tilde{\Gamma}^{\lambda}_{\beta\mu}\tilde{\Gamma}^{\beta}_{\sigma\nu}
-\tilde{\Gamma}^{\lambda}_{\beta\nu}\tilde{\Gamma}^{\beta}_{\sigma\mu}.
\end{eqnarray*}

Field equations and Ricci identities look thus \cite{Schouten}:

\begin{eqnarray*}
\tilde{R}_{\mu\nu}-\frac{1}{2}g_{\mu\nu}\tilde{R} = \kappa \tilde{T}_{\mu\nu},\ 
\tilde{R}_{\mu\nu}=\tilde{R}^{\lambda}_{. \mu\lambda\nu},
\ \tilde{R}=\tilde{R}^{\mu}_{.\mu}, \\
Q^{\mu}_{. ab}+2 v^{\mu}_{[a}Q_{b]}=\kappa S^{\mu}_{. ab},\ 
\kappa = 8 \pi G_{N} c^{-4}, \\
Q_{a}=v^{\mu}_{a}Q_{\mu},\ Q_{\mu}=Q^{\mu}_{. \mu\nu},\
[a b]=1/2(a b -b a),\  (a b)=1/2(a b +b a)\\
(\tilde{\nabla}_{\mu}\tilde{\nabla}_{\nu}
-\tilde{\nabla}_{\nu}\tilde{\nabla}_{\mu})u_{\lambda}=
-\tilde{R}^{\sigma}_{. \lambda\mu\nu}u_{\sigma}
-2 Q^{\sigma}_{. \nu\mu}\tilde{\nabla}_{\sigma}u_{\lambda}, \\
\tilde{\nabla}_{\alpha}u_{\beta}=\partial_{\alpha}u_{\beta}
-\tilde{\Gamma}^{\nu}_{\beta\alpha}u_{\nu}.
\end{eqnarray*}

Conformal or Weyl tensor $\tilde{C}_{\sigma\lambda\mu\nu}$ is defined as:

\begin{eqnarray*}
\tilde{R}_{\sigma\lambda\mu\nu}=
\frac{1}{2}(g_{\sigma\mu}\tilde{R}_{\lambda\nu}
-g_{\sigma\nu}\tilde{R}_{\lambda\mu}
-g_{\lambda\mu}\tilde{R}_{\sigma\nu}
+g_{\lambda\nu}\tilde{R}_{\sigma\mu}) \\
-\frac{1}{6}\tilde{R}(g_{\sigma\mu}g_{\lambda\nu}
-g_{\sigma\nu}g_{\lambda\mu})+\tilde{C}_{\sigma\lambda\mu\nu}.
\end{eqnarray*}

The energy-momentum tensor of the Weyssenhoff fluid is
derived by Obukhov and Korotky \cite{Obukhov}:

\begin{eqnarray}
T_{\mu\nu}=-(p-\Lambda) g_{\mu\nu}+u_{\mu}[u_{\nu}(\rho+p)
+2 u^{\alpha}\tilde{\nabla}_{\beta}S^{\beta}_{. \alpha\nu}].
\end{eqnarray}

The Ehlers-decomposition of the velocity-gradient can be written as

\begin{eqnarray}
\tilde{\nabla}_{\mu}u_{\nu}=\tilde{\omega}_{\nu\mu}
+\sigma_{\mu\nu}+\frac{1}{3}\Theta h_{\mu\nu}+u_{\mu}a_{\nu}, \\
u^{\mu}u_{\mu}=1,\ h_{\mu\nu}=g_{\mu\nu}-u_{\mu}u_{\nu},
\ a_{\mu}=u^{\nu}\tilde{\nabla}_{\nu}u_{\mu}, 
\ \Theta=\tilde{\nabla}_{\nu}u^{\nu}, \nonumber \\
\tilde{\omega}_{\mu\nu}=h^{\alpha}_{\mu}h^{\beta}_{\nu}
\tilde{\nabla}_{[\beta}u_{\alpha ]},\ 
\sigma_{\mu\nu}=h^{\alpha}_{\mu}h^{\beta}_{\nu}
\tilde{\nabla}_{(\alpha}u_{\beta )}-\frac{1}{3}\Theta h_{\mu\nu}.
\nonumber
\end{eqnarray}

The vorticity is uniquely defined by a variational principle 
(see Eq. (3.11) of \cite{Obukhov}):

\begin{eqnarray}
\tilde{\omega}_{i j} = v^{\mu}_{i}(\tilde{\nabla}_{\alpha}v^{\nu}_{j})
u^{\alpha}g_{\mu\nu},\ i,j=\hat{1},\hat{2},\hat{3}.
\end{eqnarray}

The above two formulas for vorticity agree, but a formula for vorticity 
in (5.5) of \cite{Obukhov} has a wrong sign, as well as
definitions of vorticity in \cite{Chrobok1} and \cite{Chrobok2}.
Eventually, this confusion caused some wrong terms in the derivation of
the evolution equations in \cite{Palle5}, as it is pointed out in 
\cite{Chrobok2}
and later in \cite{Brechet}.

The standard procedure leads to the evolution equations
(Frenkel condition employed $u^{\mu}Q^{\kappa}_{. \mu\nu}=0$):

\begin{eqnarray*}
\dot{F}\equiv u^{\mu}\tilde{\nabla}_{\mu} F,\ 
\tilde{\omega}^{2}=\frac{1}{2}\tilde{\omega}_{\mu\nu}
\tilde{\omega}^{\mu\nu},\ 
\sigma^{2}=\frac{1}{2}\sigma_{\mu\nu}\sigma^{\mu\nu}, \\
\dot{\Theta}=\tilde{\nabla}_{\mu}a^{\mu}+2 \tilde{\omega}^{2}
-2 \sigma ^{2}-\frac{1}{3}\Theta ^{2}-\tilde{R}_{\sigma\nu}
u^{\sigma}u^{\nu}, \\
h^{[\nu}_{\alpha}h^{\lambda ]}_{\beta}\dot{\tilde{\omega}}_{\nu\lambda}
=-\frac{2}{3}\Theta \tilde{\omega}_{\alpha\beta}
-2 \sigma_{[\alpha .}^{\ \gamma}\tilde{\omega}_{\gamma | \beta ]}
-h^{[\nu}_{\alpha}h^{\lambda ]}_{\beta}\tilde{\nabla}_{\nu}a_{\lambda}
+h^{[\nu}_{\alpha}h^{\lambda ]}_{\beta}u^{\mu}u^{\kappa}
\tilde{R}_{\kappa \lambda \mu \nu}, 
\end{eqnarray*}
\begin{eqnarray*}
h_{\nu}^{\alpha}h_{\mu}^{\beta}\dot{\sigma}_{\alpha\beta} &=&
h_{\nu}^{\alpha}h_{\mu}^{\beta}\tilde{\nabla}_{(\alpha}a_{\beta)}
-a_{\nu}a_{\mu}-\tilde{\omega}_{(\nu | \rho}\tilde{\omega}^{\rho}_{. |
\mu)}-\sigma_{\nu\alpha}\sigma_{\mu .}^{\ \alpha} \\
&-&\frac{2}{3}\Theta\sigma_{\nu\mu}-\frac{1}{3}h_{\nu\mu}[2(\tilde{\omega}^{2}
-\sigma^{2})+\tilde{\nabla}_{\alpha}a^{\alpha}]-E_{(\nu\mu)}, \\
E_{\alpha\beta}&=&\tilde{C}_{\sigma\alpha\mu\beta}u^{\sigma}u^{\mu}. 
\end{eqnarray*}

We use the following identity:

\begin{eqnarray}
^{(3)}\tilde{\nabla}_{\mu}(\dot{f})-h^{\nu}_{\mu}(^{(3)}\tilde{\nabla}_{\nu}
f)^{.}
=a_{\mu}\dot{f}+(\tilde{\omega}^{\lambda}_{. \mu}+
\sigma_{\mu .}^{\ \lambda}+\frac{1}{3}\Theta h_{\mu}^{\lambda})
^{(3)}\tilde{\nabla}_{\lambda}f,
\end{eqnarray}

and the continuity equation in matter dominated epoch:

\begin{eqnarray}
u^{\mu}\tilde{\nabla}_{\mu} \rho + \rho \tilde{\nabla}_{\mu} u^{\mu}=0
\end{eqnarray}

to derive the evolution equation for density contrast Eq.(5).

We evaluate the coefficients of the coupled evolution equations
for components in the local Lorentzian frame ${\cal D}_{a}$
with the metric of Eq.(7) (time component ${\cal D}_{\hat{0}}$ can be set 
to zero because of the relation $u^{a}{\cal D}_{a}=0$):

\begin{eqnarray}
\ddot{{\cal D}}_{\hat{i}}+b_{ij}\dot{{\cal D}}_{\hat{j}}
+a_{ik}{\cal D}_{\hat{k}}+d_{i}=0,\ i,j,k=1,2,3,
\end{eqnarray}
\begin{eqnarray*}
a_{11} &=& -\frac{1}{3}\dot{\lambda}^{2}-\frac{1}{3}\kappa \Lambda
-\frac{5}{3}Q^{2}-\frac{1}{3}\kappa \rho -\frac{5}{3}\frac{\dot{R}}{R}
\lambda\dot{\lambda} \\ &-& \frac{1}{3}\ddot{\lambda}\lambda
+\frac{5}{3}\frac{m}{R}\lambda Q-\frac{2}{3}\lambda^{2}\frac
{\dot{R}^{2}}{R^{2}}-\frac{5}{12}(\lambda \frac{m}{R})^{2}
-\frac{1}{3}\frac{\ddot{R}}{R}(-3+\lambda^{2}), \\
a_{12}&=&\frac{1}{2}\frac{m}{R}\dot{\lambda}+\frac{1}{4}\frac{\dot{R}}{R}Q
+\frac{1}{2}\lambda\frac{m}{R}\frac{\dot{R}}{R},\ a_{13}=0, \\
b_{11}&=& 2 \frac{\dot{R}}{R},\ b_{12}=-2 Q + \lambda \frac{m}{R},\ 
b_{13}=0, 
\end{eqnarray*}
\begin{eqnarray*}
a_{21} &=& -\frac{1}{2}\frac{m}{R}\dot{\lambda}
-\frac{1}{4}\frac{\dot{R}}{R}Q-\frac{1}{2}\lambda\frac{m}{R}\frac{\dot{R}}{R}, \\
a_{22} &=& -\frac{4}{3}\dot{\lambda}^{2}-\frac{1}{3}\kappa\Lambda
-\frac{5}{3}Q^{2}-\frac{1}{3}\kappa\rho
-\frac{11}{3}\frac{\dot{R}}{R}\lambda\dot{\lambda} \\
&-& \frac{1}{3}\ddot{\lambda}\lambda
+\frac{5}{3}\frac{m}{R}Q\lambda-\frac{5}{3}\lambda^{2}(\frac{\dot{R}}{R})^{2}
-\frac{5}{12}(\frac{\lambda}{m R})^{2}
-\frac{1}{3}\frac{\ddot{R}}{R}(-3+\lambda^{2}), \\
a_{23} &=& 0,\ b_{21}=2 Q-\lambda\frac{m}{R},\ 
b_{22} = 2\frac{\dot{R}}{R},\ b_{23} = 0, 
\end{eqnarray*}
\begin{eqnarray*}
a_{31}&=& 0,\ a_{32}=0, \\
a_{33}&=& -\frac{1}{3}\dot{\lambda}^{2}-\frac{1}{3}\kappa\Lambda
-\frac{2}{3}Q^{2}-\frac{1}{3}\kappa\rho
-\frac{5}{3}\frac{\dot{R}}{R}\lambda\dot{\lambda} 
- \frac{1}{3}\ddot{\lambda}\lambda \\
&+&\frac{2}{3}\frac{m}{R}\lambda Q
-\frac{2}{3}\frac{\dot{R}^{2}}{R^{2}}\lambda^{2}
-\frac{1}{6}\lambda^{2}\frac{m^{2}}{R^{2}} 
-\frac{1}{3}\frac{\ddot{R}}{R}(-3+\lambda^{2}), \\
b_{31}&=& 0,\ b_{32}=0,\ b_{33}=2\frac{\dot{R}}{R},
\end{eqnarray*}
\begin{eqnarray*}
d_{1}&=&d_{3}=0, \\
d_{2}&=& R^{2} [-12\lambda (\frac{\dot{R}}{R})^{3}+\dot{\lambda} 
\frac{\dot{R}^{2}}
{R^{2}}(-6+13\lambda^{2})+2\dot{\lambda}^{3}+\lambda^{2}(\stackrel{...}
\lambda+\lambda\frac{\stackrel{...}R}{R}) \\
&+&\dot{\lambda}(4Q^{2}+\kappa (2\Lambda-\rho)-6\frac{m}{R}Q\lambda
+\lambda (5\ddot{\lambda}+2\lambda \frac{m^{2}}{R^{2}}+
9\lambda\frac{\ddot{R}}{R})) \\
&+&\frac{\dot{R}}{R}(\ddot{\lambda}(3+7\lambda^{2})+\lambda (17\dot{\lambda}^{2}
+2\kappa \Lambda -2Q^{2}-\kappa\rho+\frac{m}{R}\lambda Q) \\
&+& \lambda \frac{\ddot{R}}{R}(3+5\lambda^{2}))], \\
for & & i \neq j:\ a_{ij}=-a_{ji},\ b_{ij}=-b_{ji}.
\end{eqnarray*}

The symmetric parts of Einstein-Cartan equations are given by:

\begin{eqnarray*}
(\hat{0}\hat{0})&:& -2\lambda^{2}\frac{\ddot{R}}{R}+\frac{\dot{R}^{2}}{R^{2}}
(3-\lambda^{2})+\frac{m^{2}}{4 R^{2}}(-4+3\lambda^{2})-2\frac{\dot{R}}{R}
\dot{\lambda}\lambda=\kappa (\rho+\Lambda)+\frac{2 m \lambda Q}{R}-Q^{2}, \\
(\hat{1}\hat{1})&:& 2(1-\lambda^{2})\frac{\ddot{R}}{R}+(1-\lambda^{2})
\frac{\dot{R}^{2}}{R^{2}}-\frac{m^{2}\lambda^{2}}{4 R^{2}}-\dot{\lambda}^{2}
-5 \frac{\dot{R}}{R}\dot{\lambda}\lambda-\ddot{\lambda}\lambda 
=\kappa \Lambda +Q^{2}-\frac{m\lambda Q}{R}, \\
(\hat{2}\hat{2})&:& 2\frac{\ddot{R}}{R}+(1-3\lambda^{2})\frac{\dot{R}^{2}}
{R^{2}}-\frac{m^{2} \lambda^{2}}{4 R^{2}}-2\frac{\dot{R}}{R}\dot{\lambda}
\lambda = \kappa \Lambda +Q^{2}-\frac{m \lambda Q}{R}, \\
(\hat{3}\hat{3})&:& (1-\lambda^{2})(2\frac{\ddot{R}}{R}+\frac{\dot{R}^{2}}
{R^{2}})-(4-\lambda^{2})\frac{m^{2}}{4 R^{2}}-5 \frac{\dot{R}}{R}
\dot{\lambda}\lambda-\dot{\lambda}^{2}-\ddot{\lambda}\lambda
=\kappa \Lambda +Q^{2}, \\
(\hat{0}\hat{1})&:& \frac{m \lambda}{R} (\lambda \frac{\dot{R}}{R}
+\frac{3}{2}\dot{\lambda})=\dot{Q}\lambda+2\dot{\lambda}Q
+3\lambda Q\frac{\dot{R}}{R}, \\
(\hat{0}\hat{2})&:& 2\lambda (\frac{\dot{R}^{2}}{R^{2}}
-\frac{\ddot{R}}{R}) = 0, \\
(\hat{1}\hat{2})&:& \frac{m}{2 R} (\dot{\lambda}+2\lambda \frac{\dot{R}}{R})=0.
\end{eqnarray*}

In the limit of small $\lambda^{2} << 1$ we combine $\hat{0}\hat{0}$
and $\hat{1}\hat{1}$ components to approximate the time gradients 
of the cosmic scale factor:

\begin{eqnarray}
\frac{\dot{R}^{2}}{R^{2}}&=&\frac{1}{3}(\kappa (\rho+\Lambda)
-Q^{2}+\frac{2 m \lambda Q}{R}+(\frac{m}{R})^{2}),  \nonumber \\
\frac{\ddot{R}}{R}&=&\frac{1}{2}(\frac{2}{3}\kappa \Lambda
-\frac{1}{3}\kappa \rho+\frac{4}{3} Q^{2}-\frac{5}{3}\frac{m\lambda Q}{R}
-\frac{1}{3} (\frac{m}{R})^{2}). 
\end{eqnarray}

The age of the Universe follows immediately:

\begin{eqnarray}
\tau_{U}(Gyr)=\frac{10}{h}\int^{1}_{10^{-3}}\frac{d a}{a}
[\Omega_{\Lambda}+\Omega_{m}a^{-3}-\frac{1}{3}Q^{2}
+\frac{2}{3}\frac{m\lambda Q}{a}+\frac{m^{2}}{3 a^{2}}]^{-1/2}, \\
\lambda=\lambda_{0}a^{-1},\ Q=Q_{0}a^{-3/2};\ 
m,\ \lambda_{0},\ Q_{0}\ evaluated\ in\ the\ unit \ H_{0}=1. \nonumber
\end{eqnarray}

Note that even a term linear in torsion $Q$ is negative
because \cite{Palle4} $m\lambda > 0 (m\lambda < 0)$ implies
$Q < 0 (Q > 0)$. 

The relation (3.31) of Ref. \cite{Obukhov}, as the equation of
motion for the angular momentum of the Zel'dovich model, is
no more valid.

Let us finally write the CDM spectrum used in numerical
evaluations \cite{Kolb}:

\begin{eqnarray}
P(\vec{k})&=&|\delta_{\vec{k}}|^{2}
=\frac{A k}{(1+\beta k+\alpha k^{1.5}+\gamma k^{2})^{2}}, \\
k=|\vec{k}|,\ 
\beta &=& 1.7(\Omega_{m}h^{2})^{-1} Mpc,\ 
\alpha=9.0(\Omega_{m}h^{2})^{-1.5} Mpc^{1.5},\ 
\gamma=1.0(\Omega_{m}h^{2})^{-2} Mpc^{2}. \nonumber
\end{eqnarray}

\section{Appendix B}

One can introduce the following general vector variable fulfilling the
Stewart-Walker lemma
${\cal D}_{\mu}(k) \equiv  R^{k}(t)\rho^{-1}h_{\mu}^{\ \nu}
\tilde{\nabla}_{\nu} \rho$.
It is easy to verify that the correct Friedmann limes of 
density contrast can be 
achieved by rescaling the scalar invariant of the vector variable

\begin{eqnarray*}
\delta \propto R^{2-k}(t)[-{\cal D}_{\mu}(k){\cal D}^{\mu}(k)]^{1/2}.
\end{eqnarray*}

We show below that our choice ($k=2$) and rescaled Ellis-Bruni choice 
($k=1$) \cite{Ellis}
give different results for geometries beyond that of Friedmann.
Thus, only our choice of the variable ($k=2$) gives a correct density
contrast without any ad hoc posterior rescaling.

Ellis-Bruni covariant vector variables are defined as \cite{Ellis}:

\begin{eqnarray}
\bar{{\cal D}}_{\mu} &\equiv & R(t)\rho^{-1}h_{\mu}^{\ \nu}
\tilde{\nabla}_{\nu} \rho, \\
\bar{{\cal L}}_{\mu} &\equiv & R(t)h_{\mu}^{\ \nu}
\tilde{\nabla}_{\nu} \Theta. \nonumber
\end{eqnarray}

The same procedure as in Appendix A results in the following
equation for a density contrast in the matter dominated
epoch:

\begin{eqnarray}
\ddot{\bar{\cal D}}_{\mu}+a_{\mu}a^{\lambda}\bar{\cal D}_{\lambda}
+u_{\mu}\dot{a^{\lambda}}\bar{\cal D}_{\lambda}
+u_{\mu}a^{\lambda}\dot{\bar{\cal D}}_{\lambda}
-\frac{1}{2}\kappa \rho \bar{\cal D}_{\mu}  \nonumber \\
+(\dot{\bar{\cal D}}_{\lambda}+\bar{\cal D}_{\nu}(\sigma^{\nu}_{. \lambda}
+\tilde{\omega}^{\nu}_{. \lambda}))(\frac{2}{3}\Theta\delta^{\lambda}_{\mu}
+u_{\mu}a^{\lambda}+\tilde{\omega}^{\lambda}_{. \mu}+\sigma^{\lambda}_{. \mu}) 
\nonumber \\
+\frac{2}{3}\Theta u_{\mu}a^{\nu}\bar{\cal D}_{\nu}
+\dot{\bar{\cal D}}_{\lambda}(\tilde{\omega}^{\lambda}_{.\mu} 
+\sigma^{\lambda}_{. \mu})+\bar{\cal D}_{\lambda}(\dot{\sigma}^{\lambda}_{.\mu}
+\dot{\tilde{\omega}}^{\lambda}_{.\mu})   \nonumber \\    
-R [2 a_{\mu}\dot{\Theta}+ ^{(3)}\tilde{\nabla}_{\mu} {\cal N}
+\Theta a_{\lambda} (\frac{2}{3}\Theta \delta^{\lambda}_{\mu}
+u_{\mu}a^{\lambda}+\tilde{\omega}^{\lambda}_{. \mu}+\sigma^{\lambda}_{. \mu}) 
\nonumber \\
+\frac{1}{3}\Theta^{2}a_{\mu}
+\Theta \dot{a_{\mu}} ] = 0 .
\end{eqnarray}

The corresponding equations in the local Lorentzian frame are:

\begin{eqnarray}
\ddot{\bar{\cal D}}_{\hat{i}}+\bar{b}_{ij}\dot{\bar{\cal D}}_{\hat{j}}
+\bar{a}_{ik}\bar{\cal D}_{\hat{k}}+\bar{d}_{i}=0,\ i,j,k=1,2,3,
\end{eqnarray}
\begin{eqnarray*}
\bar{a}_{11} &=& 2\frac{\dot{R}^{2}}{R^{2}}+\frac{\ddot{R}}{R}
-Q^{2}-\frac{1}{2}\kappa \rho 
+\frac{m}{R}\lambda Q
-(\frac{\lambda m}{2 R})^{2}, \\
\bar{a}_{12}&=&\frac{1}{2}\frac{m}{R}\dot{\lambda}-\frac{7}{4}\frac{\dot{R}}{R}Q
+\frac{3}{2}\lambda\frac{m}{R}\frac{\dot{R}}{R},\ \bar{a}_{13}=0, \\
\bar{b}_{11}&=& 4 \frac{\dot{R}}{R},\ \bar{b}_{12}=-2 Q + \lambda \frac{m}{R},\ 
\bar{b}_{13}=0, 
\end{eqnarray*}
\begin{eqnarray*}
\bar{a}_{21}&=&-\frac{1}{2}\dot{\lambda}\frac{m}{R}
+\frac{7}{4}\frac{\dot{R}}{R}Q-\frac{3}{2}\lambda\frac{\dot{R}}{R}\frac{m}{R}, \\
\bar{a}_{22}&=&\frac{\ddot{R}}{R}-\dot{\lambda}^{2}-Q^{2}
-\frac{1}{2}\kappa\rho-2\lambda\dot{\lambda}\frac{\dot{R}}{R} \\
&+& \frac{m}{R}Q\lambda-(\lambda\frac{m}{2 R})^{2}
-(\frac{\dot{R}}{R})^{2}(-2+\lambda^{2}), \\
\bar{a}_{23} &=& 0, 
\bar{b}_{21}=2 Q -\lambda \frac{m}{R},\ \bar{b}_{22}=4\frac{\dot{R}}{R},\ 
\bar{b}_{23}=0, \\
\bar{a}_{31}&=& 0,\ \bar{a}_{32}=0,\ 
\bar{a}_{33}=2\frac{\dot{R}^{2}}{R^{2}}+\frac{\ddot{R}}{R}
-\frac{1}{2}\kappa \rho, \\
\bar{b}_{31}&=& 0,\ \bar{b}_{32}=0,\
\bar{b}_{33}=4 \frac{\dot{R}}{R}, 
\end{eqnarray*}
\begin{eqnarray*}
\bar{d}_{1}&=&\bar{d}_{3}=0, \\
\bar{d}_{2}&=& R [2 \dot{\lambda}^{3}
+\lambda^{2}(\stackrel{...}\lambda+\frac{\stackrel{...}R}{R})
+(\frac{\dot{R}}{R})^{2}\dot{\lambda}(6+13\lambda^{2}) \\
&+&\dot{\lambda}(4 Q^{2}+\kappa (2\Lambda -\rho)
-6 \frac{m}{R}\lambda Q \\
&+& \lambda (5\ddot{\lambda}+2\lambda \frac{m^{2}}{R^{2}}
+9\lambda \frac{\ddot{R}}{R})) \\
&+& \frac{\dot{R}}{R}(\ddot{\lambda}(3+7 \lambda^{2})
+\lambda(17\dot{\lambda}^{2}+2\kappa \Lambda -2Q^{2} \\
&-& \kappa\rho +\frac{m}{R}\lambda Q+\lambda \frac{\ddot{R}}{R}
(3+5\lambda^{2}))] , \\
for & &\ i\neq j:\ \bar{a}_{ij}=-\bar{a}_{ji},\ 
\bar{b}_{ij}=-\bar{b}_{ji}.
\end{eqnarray*}

The authors in \cite{Ellis} equalize components of their variables
$\bar{\cal D}_{\mu}$ with a scalar density contrast $\delta$.
This is possible in the Friedmann limes when all components are
equal. However even then, the scalar quantity formed from their
variables must be ad hoc multiplied by the cosmic scale factor
to achieve the correct result:

\begin{eqnarray*}
\delta \propto R(t)[-\bar{\cal D}_{\mu}\bar{\cal D}^{\mu}]^{1/2}.
\end{eqnarray*}

Our corrected variables ${\cal D}_{\mu}$, on the contrary, give 
immediately good and correct Friedmann limes:

\begin{eqnarray}
\delta \propto [-{\cal D}_{\mu}{\cal D}^{\mu}]^{1/2}
= [-{\cal D}_{a}{\cal D}^{a}]^{1/2}
\end{eqnarray}

Let us stress that beyond Friedmannian geometry two quantities are not
equal:

\begin{eqnarray*}
R(t)[-\bar{\cal D}_{\mu}\bar{\cal D}^{\mu}]^{1/2}
\neq [-{\cal D}_{\mu}{\cal D}^{\mu}]^{1/2},
\end{eqnarray*}

hence we use throughout our paper corrected variables ${\cal D}_{\mu}$.
In Fig. 4 the reader can find comparison between two formulas
when the vorticity and the acceleration do not vanish.

\begin{figure}
\epsfig{figure=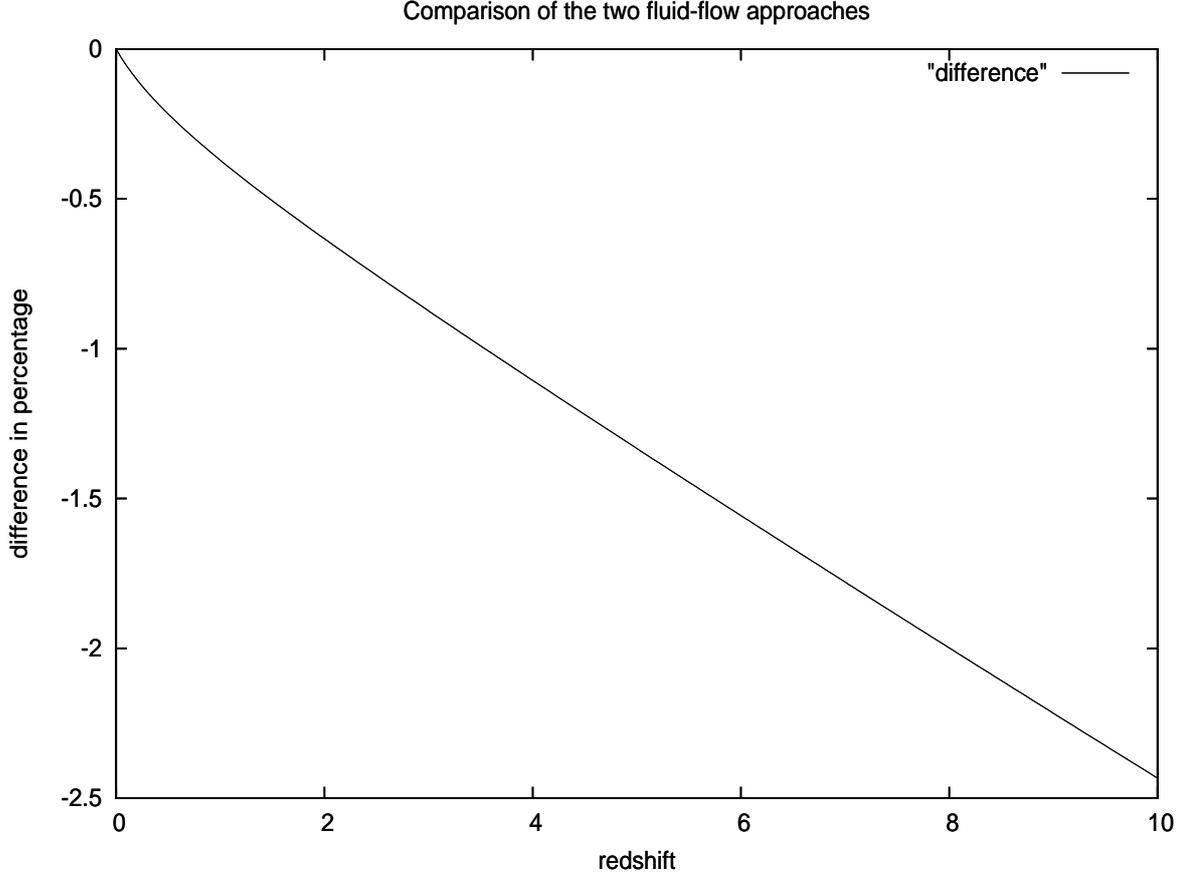, height=120 mm, width=160 mm}
\caption{Comparison between the two fluid-flow approaches for the model EC3:
$\delta \equiv [-{\cal D}_{\mu}{\cal D}^{\mu}]^{1/2}$ and
$\bar{\delta} \equiv R(t)[-\bar{\cal D}_{\mu}\bar{\cal D}^{\mu}]^{1/2}$, 
difference $\equiv (\delta-\bar{\delta})/\delta$, $\delta (z=0)=
\bar{\delta} (z=0)=1$.}
\end{figure}

\end{document}